\documentclass[a4paper,USenglish,final,draftx]{lipics}
\usepackage{amsmath}
\usepackage{amssymb}
\usepackage{amsthm}
\usepackage{stmaryrd}
\usepackage{subfig}
\usepackage{showlabels}
\usepackage{tikz}
\usepackage[sort,noadjust]{cite}
\usepgflibrary{arrows}
\usepgflibrary[arrows]
\usetikzlibrary{shapes,arrows}
\usetikzlibrary[arrows]
\usepackage{xspace}
\usepackage[size=small]{isabelle}
\usepackage{url}

\newcommand{\ACL}{ACL\xspace}
\newcommand{\COQ}{Coq\xspace}
\newcommand{\HOL}{HOL\xspace}
\newcommand{\Isabelle}{Isabelle\xspace}
\newcommand{\ISAFOR}{IsaFoR\xspace}
\newcommand{\ISAR}{Isar\xspace}

\newcommand{\PVS}{PVS\xspace}
\newcommand{\ds}[1]{{{\raisebox{2.4mm}{\rotatebox{270}{$\succ$}}}{#1}}}

\newcommand{\mult}{\prec_\textsf{mult}}
\newcommand{\multone}{\prec_\textsf{mult1}}
\newcommand{\muleq}{\preccurlyeq_\textsf{mul}}
\newcommand{\mul}{\prec_\textsf{mul}}
\newcommand{\pex}{\prec_\textsf{peak}}

\newcommand{\nil}{[\,]}
\newcommand{\set}[1]{\textsf{set~}#1}
\newcommand{\setof}[1]{\textsf{set\_of~}#1}
\newcommand{\multisetof}[1]{\textsf{multiset\_of~}#1}
\newcommand{\length}[1]{\textsf{length~}#1}

\renewcommand{\set}[1]{#1}
\renewcommand{\setof}[1]{#1}
\renewcommand{\multisetof}[1]{#1}

\newcommand{\diffs}{\mathrel{-s}}
\newcommand{\inters}{\mathbin{{\cap}s}}
\newcommand{\interm}{\mathbin{{\#}{\cap}}}
\newcommand{\ins}{\mathbin{{\in}{\#}}}
\newcommand{\notins}{\mathbin{{\notin}{\#}}}

\renewcommand{\AA}{\mathcal{A}}
\newcommand{\BB}{\mathcal{B}}
\newcommand{\lto}[1]{\stackrel{#1}{\to}}
\newcommand{\ltoe}[1]{\mathrel{\smash{\stackrel{#1}{\to}}^=}}
\newcommand{\ltoo}[1]{\stackrel{#1}{\to\!\!\!\!\!\to}}
\newcommand{\from}{\leftarrow}
\newcommand{\lfrome}[1]{\mathrel{\smash{{}^={\stackrel{#1}{\from}}}}}
\newcommand{\lfrom}[1]{\stackrel{#1}{\from}}
\newcommand{\lfromm}[1]{\stackrel{#1}{\from\!\!\!\!\!\from}}

\newcommand{\lconv}[1]{\stackrel{#1}{\leftrightarrow}}
\newcommand{\lconvv}[1]{\stackrel{#1}{\leftrightarrow\!\!\!\!\!\leftrightarrow}}
\newcommand\Qed{\qed\medskip}
\newcommand{\thy}[1]{\textsf{#1}}
\newcommand{\drop}[1]{}
\newcommand{\isa}[1]{\hol{#1}}

\title{Confluence by Decreasing Diagrams -- Formalized%
}
\author{Harald Zankl}
\authorrunning{H.~Zankl}

\affil{Institute of Computer Science, University of Innsbruck,
6020 Innsbruck, Austria%\\
}

\subjclass{
 F.3.1, %Specifying and Verifying and Reasoning about Programs,
 F.4.2 %Grammars and Other Rewriting Systems
}
\keywords{term rewriting, confluence, decreasing diagrams, formalization}

\begin{document}

\maketitle

\begin{abstract}
This paper presents a formalization of decreasing diagrams in the
theorem prover \Isabelle. It discusses mechanical proofs showing that
any locally decreasing abstract rewrite system is confluent. The
valley and the conversion version of decreasing diagrams are considered.
\end{abstract}

\section{Introduction}

Formalizing confluence criteria has a long history in $\lambda$-calculus.
Huet~\cite{H94} proved a stronger variant of the parallel moves lemma in
\COQ.
\Isabelle/\HOL was used in~\cite{N01} to prove the Church-Rosser property
of $\beta$, $\eta$, and $\beta\eta$. For $\beta$-reduction the
standard Tait/Martin-L{\"o}f proof as well as
Takahashi's proof~\cite{T95} were formalized. The first mechanically
verified proof of the Church-Rosser property of $\beta$-reduction was
done using the Boyer-Moore theorem prover~\cite{S88}. The formalization
in Twelf~\cite{P92} was used to formalize the confluence proof of a
specific higher-order rewrite system in~\cite{S06}.

Newman's lemma (for abstract rewrite systems) and Knuth and Bendix' 
critical pair theorem (for first-order rewrite systems) have been
proved in~\cite{RRAHMM02} using \ACL.
An alternative proof of the latter in \PVS, following the higher-order
structure of Huet's proof, is presented in~\cite{GAR10}.
\PVS is also used in the formalization of the lemmas of Newman and
Yokouchi in~\cite{GAR08}. Knuth and Bendix' criterion has also been
formalized in \COQ~\cite{CCFPU11} and \Isabelle/\HOL~\cite{T12}. 

Decreasing diagrams~\cite{vO94} are a complete characterization of confluence
for abstract rewrite systems whose convertibility classes are countable.
As a criterion for abstract rewrite systems, they can easily be applied for
first- and higher-order rewriting, including term rewriting and the 
$\lambda$-calculus. Furthermore, decreasing diagrams yield constructive proofs of
confluence~\cite{vO08b} (in the sense that the joining sequences can be computed
based on the divergence).
We are not aware of a (complete) formalization of decreasing diagrams in 
any theorem prover (see remarks in Section~\ref{con:main}).

In this paper we discuss a formalization of decreasing diagrams in the theorem 
prover \Isabelle/\HOL. (In the sequel we just call it \Isabelle.) 
We closely follow the proofs in~\cite{vO94,vO08a}. For alternative proofs
see~\cite{BKO98,KOV00} or~\cite{JvO09,vO12,F12c} where proof orders
play an essential role.
The main contributions of this paper are (two) mechanical proofs of 
Theorem~\ref{thm:main} in \Isabelle.

\begin{theorem}[\textnormal{\cite{vO94,vO08a}}]
\label{thm:main}
A locally decreasing abstract rewrite system is confluent.
\qed
\end{theorem}

As a consequence all definitions (lemmata) in this paper have been formalized (proved) 
in \Isabelle. The definitions from the paper are (modulo notation) identical
to the ones used in \Isabelle.
Our formalization (\thy{Decreasing\_Diagrams.thy}, available from~\cite{Z13corr})
consists of approximately 1600 lines of \Isabelle code in the \ISAR style and
contains 31 definitions and 122 lemmata.
The valley version~\cite{vO94} amounts to ca.\ 1000 lines, 22 definitions, 
and 97 lemmata while the conversion version~\cite{vO08a} has 
additional 600 lines of \Isabelle comprising 9 definitions and 25 lemmata.
Our formalization imports the theory \thy{Multiset.thy} from the \Isabelle library
and \thy{Abstract\_Rewriting.thy}~\cite{ST10} from the Archive of Formal Proofs.
We used \Isabelle 2012 and the Archive of Formal Proofs from July 30, 2012.

The remainder of this paper is organized as follows. In the next section we
recall helpful preliminaries for our formalization of~\cite{vO94},
which is described in Section~\ref{for:main}. The conversion
version of decreasing diagrams~\cite{vO08a} is the topic of Section~\ref{for:cv}.
In Section~\ref{mea:main} we highlight changes to (and omissions in) the proofs 
from~\cite{vO94,vO08a} before we conclude in Section~\ref{con:main}.
Appendix~\ref{def:main} presents the most important definitions in \Isabelle
notation.

\section{Preliminaries}
\label{pre:main}

We assume familiarity with rewriting~\cite{TeReSe} and 
decreasing diagrams~\cite{vO94}.
Basic knowledge of \Isabelle~\cite{ISABELLE} is not essential but may be helpful.

Given a relation $\to$ we write $\from$ for its inverse, $\ltoo{}$ for its
transitive closure, and $\ltoe{}$ (in pictures also $\lto{=}$) for its reflexive
closure. We write $\leftrightarrow$ for $\to$ or $\from$ and denote 
sets by $S$, $T$,~$U$,
multisets by $M$, $N$, $I$, $J$, $K$, $Q$, 
single labels by $\alpha$, $\beta$, and $\gamma$, and
lists of labels by
$\sigma$, $\tau$, $\upsilon$, $\kappa$, $\mu$, and $\rho$ (possibly primed or indexed).

\begin{table}[t]
\centering
\begin{tabular}{@{}lllll@{}}
meaning             & set & multiset & sequence/list                  & \cite{vO94} 
\\ \hline
empty               & $\{\}$          & $\{\#\}$      & $\nil$        & $\emptyset$/$\epsilon$ \\
singleton           & $\{\alpha\}$    & $\{\#\alpha\#\}$& $[\alpha]$  & $\{\alpha\}/[\alpha]/\alpha$\\
membership          & $\alpha \in S$  & $\alpha \ins M$ & --          & $\in$\\
union/concatenation & $S \cup T$      & $M + N$       & $\sigma@\tau$ & $\uplus$/$\sigma\tau$\\
intersection        & $S \cap T$      & $M \interm N$ & --            & $\cap$\\
difference          & $S - T$         & $M - N$       & --            & $-$\\
sub(multi)set       & $S \subseteq T$ & $M \leq N$    & --            & $\subseteq$\\
\hline
\end{tabular}
\caption{Predefined \Isabelle operators.}
\label{TAB:op}
\end{table}

Table~\ref{TAB:op} gives an overview of several predefined operators 
in \Isabelle for sets, multisets, and lists (sequences) where we also
incorporated the notation from~\cite{vO94} in the rightmost column. 
In the paper we will use the \Isabelle notation, but drop the $@$ for
concatenating sequences and write $\alpha$ instead of $[\alpha]$.
In addition to the operators provided by \Isabelle, we need the difference
(intersection) of a multiset with a set. Here $M \diffs S$ ($M \inters S$)
removes (keeps) all occurrences of elements in~$M$ that are in~$S$.
Sometimes it will be necessary to convert e.g.\ a multiset to a set (or a list).
In the paper we leave these conversions implicit, since no confusion can arise.
We establish the following useful equivalences:
\begin{lemma}[\textnormal{parts of \cite[Lemma~A.3]{vO94}}]
\label{lemmaA_3}
\mbox{}
\begin{enumerate}
\item
$(M + N) \diffs S = (M \diffs S) + (N \diffs S)$
\item
$(M \diffs S) \diffs T = M \diffs (S \cup T)$
\item
$M = (M \inters S) + (M \diffs S)$
\item
$(M \diffs T) \inters S = (M \inters S) \diffs T$
\end{enumerate}
\end{lemma}
\proof
By unfolding the definitions of multiset and the operators.
\Qed

\section{Formalization of Decreasing Diagrams}
\label{for:main}

We assume familiarity with the original proof of decreasing diagrams 
in~\cite{vO94}, upon which our formalization in this section is based.
Nevertheless we will recall the important definitions and lemmata. However,
we only give proofs if our proof deviates from the original argument. In
addition we state (sometimes small) key results, since an effective collection
of lemmata is crucial for completely formal proofs.

The remainder of this section is organized as follows:
Section~\ref{for:mul} describes our results on multisets. 
Section~\ref{for:dd} is dedicated to decreasingness (of sequences of labels) and
Section~\ref{for:ld} is concerned with an alternative formulation of
local decreasingness. Afterwards, 
Section~\ref{for:rew} lifts decreasingness (from labels) to diagrams.
Well-foundedness of the measure (on peaks) is proved in Section~\ref{for:mr},
where we also establish the main result.

\subsection{Multisets}
\label{for:mul}

In the sequel we assume $\prec$ to be a transitive and irreflexive binary relation.

\begin{definition}[\textnormal{\cite[Definition 2.5]{vO94}}]
\label{def:multiset} \mbox{}
\begin{enumerate}
\item
The set $\ds{\alpha}$ is the strict order ideal generated by (or \emph{down-set} of) $\alpha$, 
defined by $\ds{\alpha} = \{\beta \mid \beta \prec \alpha\}$. This is extended to 
sets $\ds{S} = \bigcup_{\alpha \in S} \ds{\alpha}$. We define $\ds{M}$ and $\ds{\sigma}$ to be the
down-set generated by the set of elements in $M$ and $\sigma$, respectively.
\item
The \emph{(standard) multiset extension} (denoted by $\mul$) of 
$\prec$ is defined by 
\[
\text{$M \mul N$ if $\exists$ $I$ $J$ $K$. $M = I + K$, $N = I + J$, 
 $\setof{K} \subseteq \ds{J}$, and $J \not= \{\#\}$}
\]
The relation $\muleq$ is obtained by removing the last condition ($J \not= \{\#\}$).
Note that $\muleq$ is the reflexive closure of $\mul$
(cf.\ Lemma~\ref{lem:multeq} in Section~\ref{mea:main}).
\end{enumerate}

\end{definition}

The following result is not mentioned in~\cite{vO94}---while 
\cite[Proposition~1.4.8(3)]{vO94t} shows a more general result---but 
turned out handy for our formalization. 
\begin{lemma}
\label{lem:ds_ds_subseteq_ds}
$\ds{(\ds{S})} \subseteq \ds{S}$
\end{lemma}
\proof
Assume $x \in \ds{(\ds{S})}$. By Definition~\ref{def:multiset} there must be a $y \in \ds{S}$
with $x \prec y$. From $y \in \ds{S}$ we obtain a $z \in S$ with
$y \prec z$. Then $x \prec z$ by transitivity of~$\prec$ and hence $x \in \ds{S}$.
\Qed

The multiset extension inherits some properties of the base relation, 
which we will implicitly use in the sequel.

\begin{lemma}
Let $\prec$ be a transitive and well-founded relation.
Then $\mul$ is transitive and well-founded,
and $\muleq$ is reflexive and transitive.
\end{lemma}
\proof
By Lemmata~\ref{lem:mult} and~\ref{lem:multeq} 
in combination with existing results in \thy{Multiset.thy}.
\Qed

We can now establish the following properties.

\begin{lemma}[\textnormal{\cite[Lemma~2.6]{vO94}}]
\label{lemma2_6} \mbox{}
\begin{enumerate}
\item $\ds{(S \cup T)} = \ds{S} \cup \ds{T}$ and
      $\ds{(\sigma\tau)} = \ds{\sigma} \cup \ds{\tau}$ and
      $\ds{(M \diffs S)} \supseteq \ds{M} \diffs \ds{S}$
\item $M \leq N \Rightarrow M \muleq N \Rightarrow \ds{M} \subseteq \ds{N}$ 
\item $M \muleq N \Rightarrow \exists\ I\ J\ K.\ M = I + K \land N = I + J 
 \land \setof{K} \subseteq \ds{J} \land J \interm K = \{\#\} $
\item $N \not= \{\#\} \land \setof{M} \subseteq \ds{N} \Rightarrow M \mul N$
\item $M \muleq N \Rightarrow M \diffs \ds{S} \muleq N \diffs \ds{S}$
\item $M \muleq N \Leftrightarrow Q+M \muleq Q+N$
\item $\setof{Q} \subseteq \ds{N} - \ds{M} \land M \muleq N \Rightarrow Q+M \muleq N$
\item $S \subseteq T \Rightarrow M \diffs T \muleq M \diffs S$
\item $M \mul N \Rightarrow Q+M \mul Q+N$
\end{enumerate}
\end{lemma}

Note that statements (5) and (6) slightly differ
from~\cite[Lemma~2.6]{vO94}(5,6), but are easier to apply. 
The (easy) the statements of (8) and (9) are not mentioned in~\cite{vO94},
which we required for \cite[Lemmata~3.5 and~3.6]{vO94}.

\subsection{Decreasingness}
\label{for:dd}

We define the \emph{lexicographic maximum} measure, which maps lists to
multisets, inductively.

\begin{definition}[\textnormal{\cite[Definition~3.2]{vO94}}]
\label{def:lexmax}\mbox{}
\begin{itemize}
\item $|\nil| = \{\#\}$
\item $|\alpha\sigma| = \{\#\alpha\#\} + (|\sigma| \diffs \ds{\alpha}) $
\end{itemize}
\end{definition}

The next lemma establishes properties of the lexicographic maximum measure.

\begin{lemma}[\textnormal{\cite[Lemma~3.2]{vO94}}]
\label{lemma3_2} \mbox{}
\begin{enumerate}
\item $\ds{|\sigma|} = \ds{\sigma}$
\item $\ds{|\sigma\tau|} = |\sigma| + (|\tau| \diffs \ds{\sigma})$
\end{enumerate}
\end{lemma}
\proof
\mbox{}
\begin{enumerate}
\item
By induction on~$\sigma$.
The base case is trivial. Using Lemma~\ref{lemma2_6}(1) the inductive step amounts to
\(
\ds{\alpha} \cup \ds{(|\sigma| \diffs \ds{\alpha})} = \ds{\alpha} \cup \ds{\sigma}.
\)
The inclusion from left to right follows from the induction hypothesis. For the
inclusion from right to left we proceed by case analysis.
If $x \in \ds{\alpha}$ then the result immediately follows.
If $x \notin \ds{\alpha}$ then $x \in \ds{\sigma}$ and from the
induction hypothesis $x \in \ds{|\sigma|}$. Furthermore $x \notin \ds{\alpha}$
using Lemma~\ref{lem:ds_ds_subseteq_ds} also yields $x \notin \ds{(\ds{\alpha})}$.
Hence $x \in \ds{|\sigma|} \diffs \ds{(\ds{\alpha})}$ and from Lemma~\ref{lemma2_6}(1)
we obtain $x \in \ds{(|\sigma| \diffs \ds{\alpha})}$, from which the result follows.
\item
By induction on~$\sigma$, see~\cite{vO94}.
\qed
\end{enumerate}

\emph{Decreasingness} is defined on quadruples (of sequences of labels).

\begin{definition}[\textnormal{\cite[Definition~3.3]{vO94} for labels}]
\label{def:decreasing}
The quadruple of labels $(\tau,\sigma,\sigma',\tau')$ is \emph{decreasing} (D) if 
\(
|\sigma\tau'| \muleq |\tau| + |\sigma| \text{ and } 
|\tau\sigma'| \muleq |\tau| + |\sigma|.
\)
For a visualization see Figure~\ref{fig:dd}.%
\footnote{Although the results in Sections~\ref{for:dd} and~\ref{for:ld} are on
labels only for visualization we already use diagrams.
}
\end{definition}

\begin{figure}
\begin{center}
\subfloat[Decreasing diagram.\label{fig:dd}]{
\qquad
\begin{tikzpicture}[scale=1.7]
\footnotesize
\node at (0,1) (s) {};
\node at (1,1) (s1) {};
\node at (0,0) (s2) {};
\node at (1,0) (s3) {};
\node at (0.5,0.5) {D};

\path[->>]  (s) edge node[above] {$\tau$} (s1);
\path[->>]  (s) edge node[left] {$\sigma$} (s2);
\path[->>] (s1) edge node[right] {$\sigma'$} (s3);
\path[->>] (s2) edge node[below] {$\tau'$} (s3);
\end{tikzpicture}
\qquad
}
\qquad \qquad
\subfloat[Locally decreasing diagram.\label{fig:ld}]{
\qquad
\begin{tikzpicture}[scale=1.7]
\footnotesize
\node at (0,1) (s) {};
\node at (1,1) (s1) {};
\node at (0,0) (s2) {};
\node at (1,0) (s3) {};
\node at (0.5,0.5) {LD};

\path[->]  (s)  edge node[above] {$\beta$} (s1);
\path[->]  (s)  edge node[left] {$\alpha$} (s2);
\path[->>] (s1) edge node[right] {$\sigma'$} (s3);
\path[->>] (s2) edge node[below] {$\tau'$} (s3);
\end{tikzpicture}
\qquad
}
\end{center}
\caption{Diagrams.}
\end{figure}

We write D into a diagram to indicate that its labels are decreasing.

Decreasingness can also be stated differently.

\begin{lemma}[\textnormal{\cite[Definition~3.3]{vO94}}]
\label{lem:decreasing}
The following two statements are equivalent:
\begin{enumerate}
\item
$|\sigma\tau'| \muleq |\tau| + |\sigma|$ and  
$|\tau\sigma'| \muleq |\tau| + |\sigma|$
\item
$|\tau'| \diffs \ds{\sigma} \muleq |\tau|$ and
$|\sigma'| \diffs \ds{\tau} \muleq |\sigma|$
\end{enumerate}
\end{lemma}
\proof
By Lemma~\ref{lemma3_2}(2) and Lemma~\ref{lemma2_6}(6).
\Qed

We have followed the (involved) proofs in~\cite{vO94} that pasting preserves
decreasingness (Lemma~\ref{lemma3_5}) and 
that pasting is hypothesis decreasing (Lemma~\ref{lemma3_6}) 
without big changes.

\begin{figure}
\subfloat[Lemma~\ref{lemma3_5}.\label{fig:lemma3_5}]{
\begin{tikzpicture}[scale=1.7]
\footnotesize
\node at (0,1) (s) {};
\node at (1,1) (s1) {};
\node at (0,0) (s2) {};
\node at (1,0) (s3) {};
\node at (0.5,0.5)  {D};

\node at (2,1) (t1) {};
\node at (2,0) (t3) {};
\node at (1.5,0.5)  {D};

\path[->>]  (s) edge node[above] {$\tau$} (s1);
\path[->>]  (s) edge node[left] {$\sigma$} (s2);
\path[->>] (s1) edge node[left] {$\sigma'$} (s3);
\path[->>] (s2) edge node[below] {$\tau'$} (s3);
\path[->>] (s1) edge node[above] {$\upsilon$} (t1);
\path[->>] (s3) edge node[below] {$\upsilon'$} (t3);
\path[->>] (t1) edge node[right] {$\sigma''$} (t3);
\end{tikzpicture}
\raisebox{10mm}{$\Rightarrow$}
\begin{tikzpicture}[scale=1.7]
\footnotesize
\node at (0,1) (s) {};
\node at (2,1) (s1) {};
\node at (0,0) (s2) {};
\node at (2,0) (s3) {};
\node at (1.0,0.5)  {D};

\path[->>]  (s) edge node[above] {$\tau\upsilon$} (s1);
\path[->>]  (s) edge node[left] {$\sigma$} (s2);
\path[->>] (s1) edge node[right] {$\sigma''$} (s3);
\path[->>] (s2) edge node[below] {$\tau'\upsilon'$} (s3);
\end{tikzpicture}
}
\hfill
\subfloat[Lemma~\ref{lemma3_6}.\label{fig:lemma3_6}]{
\begin{tikzpicture}[scale=1.7]
\footnotesize
\node at (0,1) (s) {};
\node at (1,1) (s1) {};
\node at (0,0) (s2) {};
\node at (1,0) (s3) {};
\node at (0.5,0.5) {D};

\path[->>]  (s) edge node[above] {$\tau$} (s1);
\path[->>]  (s) edge node[left] {$\sigma$} (s2);
\path[->>] (s1) edge node[right] {$\sigma'$} (s3);
\path[->>] (s2) edge node[below] {$\tau'$} (s3);
\path[->>] (s1) edge node[above] {$\upsilon$} (t1);
\end{tikzpicture}
}
\caption{Pasting preserves decreasingness and is hypothesis decreasing.}
\end{figure}
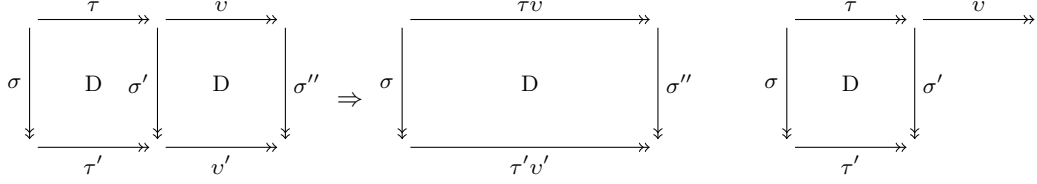

\begin{lemma}[\textnormal{\cite[Lemma~3.5]{vO94} for labels}]
\label{lemma3_5}
If $(\tau,\sigma,\sigma',\tau')$ and $(\upsilon,\sigma',\sigma'', \upsilon')$ are decreasing, 
then $(\tau\upsilon,\sigma,\sigma'',\tau'\upsilon')$ is decreasing (see Figure~\ref{fig:lemma3_5}).
\end{lemma}
\proof
As in~\cite{vO94} but we show
\(
{(|\upsilon'| \diffs \ds{\sigma\tau'}) \diffs \ds{\tau} 
 \muleq (|\upsilon'| \diffs \ds{\sigma'}) \diffs \ds{\tau}}
\)
(instead of $\subseteq$)
where we needed Lemma~\ref{lemma2_6}(8)
(in the 
last sequence in \cite[Proof of Lemma~3.5]{vO94}).
\Qed

\begin{lemma}[\textnormal{\cite[Lemma~3.6]{vO94} for labels}]
\label{lemma3_6}
If $\tau$ is non-empty and we have 
that $(\tau,\sigma,\sigma',\tau')$ is decreasing 
(see Figure~\ref{fig:lemma3_6})
then $|\sigma'| + |\upsilon| \mul |\sigma| + |\tau\upsilon|$.
\end{lemma}
\proof
As in~\cite{vO94} using Lemma~\ref{lemma2_6}(9) in the second step.
\Qed

\subsection{Local Decreasingness}
\label{for:ld}

Labels 
$(\beta,\alpha,\sigma',\tau')$
are \emph{locally decreasing} (LD) if they are
decreasing and both 
$\alpha$ and $\beta$
consist of exactly one label (see
Figure~\ref{fig:ld}). 
Now, LD can also be formulated differently:

\begin{figure}
\begin{center}
\subfloat[\label{fig:ld'}Alternative formulation of local decreasingness.]{
\qquad \qquad
\begin{tikzpicture}[scale=1.0]
\footnotesize
\node at (0,3) (s) {};
\node at (3,3) (s1) {};
\node at (0,0) (s2) {};
\node at (3,0) (s3) {};
\node at (3,2) (t1) {};
\node at (3,1) (t2) {};
\node at (1,0) (u1) {};
\node at (2,0) (u2) {};
\node at (1.5,1.5)  {LD};

\path[->]  (s) edge node[above] {$\beta$} (s1);
\path[->]  (s) edge node[left] {$\alpha$} (s2);
\path[->>] (s1) edge node[right] {$\ds{\beta}$} (t1);
\path[->] (t1) edge node[left] {$=$} node[right] {$\alpha$} (t2);
\path[->>] (t2) edge node[right] {$\ds{\alpha\beta}$} (s3);
\path[->>] (s2) edge node[below] {$\ds{\alpha}$} (u1);
\path[->] (u1) edge node[above] {$=$} node[below] {$\beta$} (u2);
\path[->>] (u2) edge node[below] {$\ds{\alpha\beta}$} (s3);
\end{tikzpicture}
\qquad \qquad
}
\subfloat[\label{fig:ld2}Giving names to the joining sequences.]{
\qquad \qquad 
\begin{tikzpicture}[scale=1.0]
\footnotesize
\node at (0,3) (s) {};
\node at (3,3) (s1) {};
\node at (0,0) (s2) {};
\node at (3,0) (s3) {};
\node at (3,2) (t1) {};
\node at (3,1) (t2) {};
\node at (1,0) (u1) {};
\node at (2,0) (u2) {};

\path[->]  (s) edge node[above] {$\beta$} (s1);
\path[->]  (s) edge node[left] {$\alpha$} (s2);
\path[->>] (s1) edge node[right] {$\sigma_1$} (t1);
\path[->] (t1) edge node[left] {$=$} node[right] {$\sigma_2$} (t2);
\path[->>] (t2) edge node[right] {$\sigma_3$} (s3);
\path[->>] (s2) edge node[below] {$\tau_1$} (u1);
\path[->] (u1) edge node[above] {$=$} node[below] {$\tau_2$} (u2);
\path[->>] (u2) edge node[below] {$\tau_3$} (s3);
\end{tikzpicture}
\qquad \qquad
}
\end{center}
\caption{Local diagrams.}
\end{figure}

\begin{lemma}[\textnormal{\cite[Prop.~3.4]{vO94}}]
\label{proposition3_4}
The form of locally decreasing labels is specified in Figure~\ref{fig:ld'}.
\end{lemma}

To show Lemma~\ref{proposition3_4} we give names to the joining sequences
as in Figure~\ref{fig:ld2}.
Then the condition of Figure~\ref{fig:ld'} can be expressed as:%
\footnote{Here \length computes the length of a list.}
\begin{align*}
\text{LD}' := {} &\set{\sigma_1} \subseteq \ds{\beta} \land
\length{\sigma_2} \leq 1 \land \set{\sigma_2} \subseteq \{\alpha\} \land
\set{\sigma_3} \subseteq \ds{\alpha\beta} \land {}
\\
&\set{\tau_1} \subseteq \ds{\alpha} \land
\length{\tau_2} \leq 1 \land \set{\tau_2} \subseteq \{\beta\} \land
\set{\tau_3} \subseteq \ds{\alpha\beta} 
\end{align*}

Local decreasingness of the labels in the diagram of Figure~\ref{fig:ld'}
(using Lemma~\ref{lem:decreasing}) yields
\begin{align*}
\text{LD} := {} &
      |\sigma' | \diffs \ds{\beta}  \muleq |\alpha| \land
      |\tau'| \diffs \ds{\alpha} \muleq |\beta| 
\end{align*}
Hence Lemma~\ref{proposition3_4} states that 
\(
\text{LD}' \text{ if and only if } \text{LD}
\).
This means that 
\begin{itemize}
\item[(i)] if a local diagram satisfies the conditions in Figure~\ref{fig:ld'}, 
i.e.\ $\text{LD}'$, then it is decreasing and 
\item[(ii)] local decreasingness implies that the
joining sequences $\tau'$ and $\sigma'$ in Figure~\ref{fig:ld} can be decomposed into
$\tau_1\tau_2\tau_3$ and $\sigma_1\sigma_2\sigma_3$ such that
the properties of the local diagram in Figure~\ref{fig:ld'}, i.e.\ $\text{LD}'$,
are satisfied.
\end{itemize}

Lemma~\ref{proposition3_4_step} will be the key result for (i),
but first we establish a useful lemma.
\begin{lemma}
\label{proposition3_4_aux} \mbox{}
$|\sigma| \leq \multisetof \sigma$
\end{lemma}
\proof 
By induction on $\sigma$. The base case is trivial. The step case
amounts to
\[
|\alpha\sigma| 
~=~ \{\#\alpha\#\} + (|\sigma| \diffs \ds{\alpha})
~\leq~ \{\#\alpha\#\} + (\sigma \diffs \ds{\alpha})
~\leq~ \multisetof{\alpha\sigma}
\]
using Definition~\ref{def:lexmax} in the first step and the induction hypothesis
in the second step.
\Qed

In the sequel we will view $|\sigma|$ and $\sigma$ as sets and use
$|\sigma| \subseteq \sigma$.
Now we can prove the following key result to establish (i).
\begin{lemma}
\label{proposition3_4_step}
$\sigma_1 \subseteq \ds{\beta} 
\land \length{\sigma_2} \leq 1 \land \sigma_2 \subseteq \{\alpha\}
\land \sigma_3 \subseteq \ds{\alpha\beta}
 \Rightarrow |\sigma_1\sigma_2\sigma_3| \diffs \ds{\beta} \muleq |\alpha|$
\end{lemma}

\proof
We show
\begin{align*}
\tag{$\star$}\!\!
(|\sigma_1| \diffs \ds{\beta})+ 
 ((|\sigma_2| \diffs \ds{\sigma_1}) \diffs \ds{\beta})+
 (((|\sigma_3| \diffs \ds{\sigma_2}) \diffs \ds{\sigma_1}) \diffs \ds{\beta})
\muleq \{\#\alpha\#\}
\end{align*}
which is equivalent to the conclusion by Lemmata~\ref{lemma3_2}(2),
\ref{lemmaA_3}(1) and Definition~\ref{def:lexmax}.
The hypothesis contains $\set{\sigma_1} \subseteq \ds{\beta}$, which together
with Lemma~\ref{proposition3_4_aux} yields 
$|\sigma_1| \subseteq \ds{\beta}$ and hence
\begin{align*}
\tag{1}
|\sigma_1| \diffs \ds{\beta} = \{\#\}
\end{align*}
Similarly from $\sigma_3 \subseteq \ds{\alpha\beta}$ we get
$|\sigma_3| \diffs (\ds{\alpha} \cup \ds{\beta}) = \{\#\}$ and hence
\begin{align*}
\tag{3}
|\sigma_3| \diffs (\ds{\sigma_2} \cup \ds{\sigma_1} \cup \ds{\alpha} \cup \ds{\beta}) = \{\#\}
\end{align*}
Using $\length{\sigma_2} \leq 1 \land \sigma_2 \subseteq \{\alpha\}$ from the
hypothesis we have two cases to consider for~$\sigma_2$.
\begin{itemize}
\item
If $\sigma_2 = \nil$ then 
\begin{align*}
\tag{2}
(|\sigma_2| \diffs \ds{\sigma_1}) \diffs \ds{\beta} = \{\#\}
\end{align*}
and from (3) we have
\begin{align*}
\tag{3'}
((|\sigma_3| \diffs \ds{\sigma_2}) \diffs \ds{\sigma_1}) \diffs \ds{\beta} \muleq \{\#\alpha\#\}
\end{align*}
using Lemma~\ref{lemmaA_3}(2).
Then $(\star)$ follows immediately from (1), (2), and (3').
\item
If $\sigma_2 = [\alpha]$ then we get (2')
\begin{align*}
(|\sigma_2| \diffs \ds{\sigma_1}) \diffs \ds{\beta} &=
|\sigma_2| \diffs (\ds{\sigma_1} \cup \ds{\beta}) 
& \text{Lemma~\ref{lemmaA_3}(2)}
\\
&= \{\#\alpha\#\} \diffs (\ds{\sigma_1} \cup \ds{\beta})
&\text{$\sigma_2 = [\alpha]$ with Definition~\ref{def:lexmax}}
\\
&\muleq \{\#\alpha\#\}
&\text{Lemma~\ref{lemma2_6}(8)}
\end{align*}
and (because $\ds{\sigma_2} = \ds{\alpha}$), similar as in the other case from (3) we get
\begin{align*}
\tag{3''}
((|\sigma_3| \diffs \ds{\sigma_2}) \diffs \ds{\sigma_1}) \diffs \ds{\beta} = \{\#\}
\end{align*}
From (1), (2'), and (3'') we conclude ($\star$).
\qed
\end{itemize}

Next we prepare for the key lemma to establish (ii), i.e.,
Lemma~\ref{proposition3_4_inv_step}, after establishing useful 
intermediate results. Note that Lemma~\ref{proposition3_4_inv_aux}(2)
can be seen as an inverse of Lemma~\ref{proposition3_4_aux}.
\begin{lemma}
\label{proposition3_4_inv_aux} \mbox{}
\begin{enumerate}
\item $\alpha \ins |\sigma| \Rightarrow 
 \exists \sigma_1\sigma_3.\ \sigma=\sigma_1\alpha\sigma_3 \land \alpha \notin \ds{\sigma_1}$
\item $\setof{|\sigma|} \subseteq \ds{S} \Rightarrow \set{\sigma} \subseteq \ds{S}$
\item $S \subseteq \ds{T} \Rightarrow \ds{S} \subseteq \ds{T}$
\end{enumerate}
\end{lemma}
\proof\mbox{}
\begin{enumerate}
\item By induction on $\sigma$. The base case is trivial. In the step case
we can assume that $\alpha \ins |\beta\sigma|$. We proceed by case analysis.
\begin{itemize}
\item
If $\alpha = \beta$ then we are done with $\sigma_1 = \nil$ and $\sigma_3 = \sigma$.
\item
In the other case we have $\alpha \ins |\sigma|$ and $\alpha \notin \ds{\beta}$
from Definition~\ref{def:lexmax}.
The induction hypothesis yields $\sigma_1'$ and $\sigma_3'$
with $\sigma = \sigma_1'\alpha\sigma_3'$ such that $\alpha \notin \ds{\sigma_1'}$.
Because $\alpha \notin \ds{\beta}$ we can conclude with $\sigma_1 = \beta\sigma_1'$ and
$\sigma_3 = \sigma_3'$ using Lemma~\ref{lemma2_6}(1).
\end{itemize}
\item
Assume $\alpha \in \set{\sigma}$. If $\alpha \ins |\sigma|$ then we are done 
by the hypothesis. In the other case there must be a $\beta \in |\sigma|$ 
(easy induction on $\sigma$) with $\alpha \prec \beta$. From the hypothesis we
get that $\beta \in \ds{S}$ and by transitivity also $\alpha \in \ds{S}$, which
finishes the proof.
\item
By monotonicity of $\ds{}$ (\cite[Proposition~1.4.8(2)]{vO94t}) the assumption yields
$\ds{S} \subseteq \ds{(\ds{T})}$. Lemma~\ref{lem:ds_ds_subseteq_ds} finishes the proof.
\qed
\end{enumerate}

With Lemma~\ref{proposition3_4_inv_aux} we can now prove the following key result
to establish (ii):
\begin{lemma}
\label{proposition3_4_inv_step}
$|\sigma'| \diffs \ds{\beta} \muleq \{\#\alpha\#\}
 \Rightarrow
\exists \sigma_1 \sigma_2 \sigma_3.\ \sigma' = \sigma_1\sigma_2\sigma_3 
 \land \setof{\sigma_1} \subseteq \ds{\beta} 
 \land {\length{\sigma_2} \leq 1} \land \setof{\sigma_2} \subseteq \{\alpha\} 
 \land \setof{\sigma_3} \subseteq \ds{\alpha\beta}$
\end{lemma}
\proof
To show the result we perform a case analysis.
\begin{itemize}
\item If $\alpha \ins |\sigma'| \diffs \ds{\beta}$ then 
Lemma~\ref{proposition3_4_inv_aux}(1) yields
$\sigma_1$ and $\sigma_3$ with $\sigma'=\sigma_1\alpha\sigma_3$ and $\alpha \notin \ds{\sigma_1}$.
Hence from the hypothesis and Lemma~\ref{lemma3_2}(2) we get
\[
 (|\sigma_1| \diffs \ds{\beta}) + \{\#\alpha\#\} 
  + (((|\sigma_3| \diffs \ds{\alpha}) \diffs \ds{\sigma_1}) \diffs \ds{\beta}) \muleq \{\#\alpha\#\} 
\]
and since 
$\alpha \notin \ds{\sigma_1}$ and $\alpha \notin \ds{\beta}$
it follows that
\[
|\sigma_1| \diffs \ds{\beta} = \{\#\} \text{ and }
 ((|\sigma_3| \diffs \ds{\alpha}) \diffs \ds{\sigma_1}) \diffs \ds{\beta} = \{\#\}
\]
Now, Lemma~\ref{lemmaA_3}(2) yields
\[|\sigma_1| \subseteq \ds{\beta} \text{ and } 
  |\sigma_3| \subseteq \ds{\alpha} \cup \ds{\sigma_1} \cup \ds{\beta}
\]
and from Lemma~\ref{proposition3_4_inv_aux}(2) we get
\[
 \set{\sigma_1} \subseteq \ds{\beta} \text{ and }
  \set{\sigma_3} \subseteq \ds{\alpha} \cup \ds{\sigma_1} \cup \ds{\beta}
\]
The latter simplifies to 
\(
  \set{\sigma_3} \subseteq \ds{\alpha\beta}
\)
using $\ds{\sigma_1} \subseteq \ds{\beta}$ (from Lemma~\ref{proposition3_4_inv_aux}(3))
and Lemma~\ref{lemma2_6}(1).
Hence in this case the result follows with $\sigma_2 = [\alpha]$.
\item If $\alpha \notins |\sigma'| \diffs \ds{\beta}$
\begin{align*}
 \Rightarrow {}&|\sigma'| \diffs \ds{\beta} \subseteq \ds{\alpha}
&\text{hypothesis}
\\
\Rightarrow {}&
|\sigma'| \subseteq \ds{\alpha\beta}
&\text{Lemma~\ref{lemma2_6}(1)}
\\
\Rightarrow {}&
\sigma' \subseteq \ds{\alpha\beta}
&\text{Lemma~\ref{proposition3_4_inv_aux}(2)}
\end{align*}
In this case the result follows with empty $\sigma_1$, empty $\sigma_2$, and
$\sigma' = \sigma_3$.
\qed
\end{itemize}

Now Lemma~\ref{proposition3_4} follows from
Lemma~\ref{proposition3_4_step} ($\text{LD}' \Rightarrow \text{LD}$) and
Lemma~\ref{proposition3_4_inv_step} ($\text{LD} \Rightarrow \text{LD}'$).

\subsection{Labeled Rewriting}
\label{for:rew}

So far we have only considered sequences of labels. However, for the main
result (Section~\ref{for:mr}) we need labeled rewriting.
Hence this section sketches how we formalized \emph{labeled} (abstract) rewriting
before lifting the results from Section~\ref{for:dd} from labels to labeled rewriting
(a step which is left implicit in~\cite{vO94}).
In the theory \thy{Abstract\_Rewriting.thy} an \emph{abstract rewrite system}
(ARS) is a set of pairs of objects of the same type, i.e., a binary relation. 
Confluence is also defined in \thy{Abstract\_Rewriting.thy}, but the theory
does not provide support for \emph{labeled} abstract rewrite systems.
In the sequel we write $\AA$ ($\BB$) for (labeled) ARSs.
A \emph{labeled ARS}~$\BB$ is a ternary relation.
We call $(a,\alpha,b) \in \BB$ a \emph{(labeled rewrite) step} and write $a \lto{\alpha} b$.
Next we define \emph{(labeled rewrite) sequences}
inductively, i.e., for each object $a$ there is the empty sequence  $a \ltoo{\nil} a$
and if $a \lto{\alpha} b$ is a 
step and $b \ltoo{\sigma} c$ is a sequence then $a \ltoo{\alpha\sigma} c$ is a sequence.

\begin{example}
\label{ex:seq}
Let $\BB$ be the labeled ARS 
$\{(a,\alpha,b), (b,\beta,c)\}$.
Then $a \lto{\alpha} b \lto{\beta} c$ (or $a \ltoo{\alpha\beta} c$) is a sequence in $\BB$. 
The empty sequence $a \ltoo{[]} a$ we also write as $a$.
\end{example}

We prove useful properties for sequences, i.e.,
that chopping off a segment of a sequence again yields a sequence and that two
sequences can be concatenated (provided the last element of the first sequence 
coincides with the first element of the second sequence).
\begin{lemma}
\label{lem:seq}
Let
$a_1 \lto{\alpha_1} \cdots \lto{\alpha_{n-1}} a_n$ and 
$b_1 \lto{\beta_1} \cdots \lto{\beta_{m-1}} b_m$
be sequences.
\begin{enumerate}
\item
Then
$a_1 \lto{\alpha_1} \cdots \lto{\alpha_{i-1}} a_i$ and
$a_i \lto{\alpha_i} \cdots \lto{\alpha_{n-1}} a_n$ are sequences
for any $1\leqslant i\leqslant n$.
\item
If $a_n = b_1$ then
$a_1 \lto{\alpha_1} \cdots \lto{\alpha_{n-1}} a_n = b_1 \lto{\beta_1} \cdots \lto{\beta_{m-1}} b_m$
is a sequence.
\end{enumerate}
\end{lemma}
\proof
By induction on $a_1 \lto{\alpha_1} \cdots \lto{\alpha_{n-1}} a_n$.
\Qed

As a next step we introduce diagrams. 
\begin{definition}
\label{def:diagram}
A \emph{diagram} is a quadruple of 
sequences $(\ltoo{\tau},\ltoo{\sigma},\ltoo{\sigma'},\ltoo{\tau'})$ such that the start and endpoints
of the sequences satisfy the picture in Figure~\ref{fig:dd}. A diagram is called
\emph{decreasing} if its labels are.
\end{definition}

We lift Lemma~\ref{lemma3_5} from labels to diagrams.

\begin{lemma}[\textnormal{\cite[Lemma~3.5]{vO94} for decreasing diagrams}]
\label{lemma3_5_DD}
Pasting two decreasing diagrams yields a decreasing diagram.
For a picture see Figure~\ref{fig:lemma3_5}.
\end{lemma}
\proof
With the help of Lemma~\ref{lem:seq}(2) we show that pasting two diagrams again
yields a diagram. That pasting preserves decreasingness follows from
Lemma~\ref{lemma3_5}.
\qed

\subsection{Main Result}
\label{for:mr}

We establish that if all local peaks of a labeled ARS~$\BB$ are decreasing
then all peaks of~$\BB$ are decreasing, following the structure of the proof of
\cite[Theorem~3.7]{vO94}. (Changes are discussed in Section~\ref{mea:main}).
Note that only here we need that $\prec$ is well-founded, from which irreflexivity
immediately follows (to satisfy our global assumption from
Section~\ref{pre:main}).
First we introduce (local) peaks.

\begin{definition}
\label{def:peak}
A \emph{peak} $(\ltoo{\tau},\ltoo{\sigma})$ is a pair of labeled rewrite
sequences which originate from the same object. A \emph{local peak} is a
peak where the sequences consist of a single step.
\end{definition}

To prove the main result we introduce a measure on \emph{peaks}
(actually on pairs of sequences).
\begin{definition}
\label{def:measure}
Let $|(\ltoo{\tau},\ltoo{\sigma})| := $ $|\tau| + |\sigma|$.
Then we can lift $\prec$ as a relation on labels to a relation on pairs of 
sequences~$\pex$, i.e., 
\(
\text{$(\ltoo{\tau},\ltoo{\sigma}) \pex (\ltoo{\tau'},\ltoo{\sigma'})$ 
if $|(\ltoo{\tau},\ltoo{\sigma})| \mul |(\ltoo{\tau'},\ltoo{\sigma'})|$}
\).
\end{definition}

For proofs of induction we establish that $\pex$ is well-founded.

\begin{lemma}
\label{lem:wf}
Let $\prec$ be well-founded. Then $\pex$ is well-founded.
\end{lemma}
\proof
From~\cite{DM79} we get that $\mul$ is well-founded (this proof is contained
in \thy{Multiset.thy}).
We proceed by contraposition. Assume the measure on peaks is not well-founded.
Then we obtain an infinite sequence
\(
\cdots \pex (\tau_2,\sigma_2) \pex (\tau_1,\sigma_1)
\)
which entails an infinite sequence on multisets
\(
\cdots \mul |\tau_2| + |\sigma_2| \mul |\tau_1| + |\sigma_1|
\)
showing the result.
\Qed

\begin{definition}
\label{def:pd}
A peak $(\ltoo{\tau},\ltoo{\sigma})$ in a labeled ARS is \emph{decreasing} if it can
be completed into a decreasing diagram, i.e., there are $\ltoo{\sigma'}$ and $\ltoo{\tau'}$
such that the conditions of Figure~\ref{fig:dd} are satisfied.
A peak is \emph{locally decreasing}, if it is decreasing and a local peak.
\end{definition}

\begin{figure}
\centering
\subfloat[\label{fig:main}Local decreasingness implies decreasingness.]{
\begin{tikzpicture}[scale=1.7]
\footnotesize
\node at (0,0) (s) {};
\node at (1,0) (s1) {};
\node at (3,0) (s2) {};

\node at (0,-1) (t) {};
\node at (1,-1) (t1) {};
\node at (3,-1) (t2) {};

\node at (0,-2) (u) {};
\node at (3,-2) (u1) {};
\path[->]  (s) edge node[above] {$\beta$} (s1);
\path[->>] (s1) edge node[above] {$\upsilon$} (s2);
\path[->>]  (t) edge node[below] {$\mu$} (t1);
\path[->>] (t1) edge node[below] {$\upsilon'$} (t2);
\path[->>] (u) edge node[below] {$\tau''$} (u1);
\node[above of=s1,yshift=-6mm] {$\tau$};
\node[left of=t,xshift=6mm] {$\sigma$};
\node[below of=t1,yshift=7mm] {$\tau'$};
\node[right of=t2,xshift=-6mm]{$\sigma'$};

\node at (0.5,-0.5) (D1) {D};
\node at (2.0,-0.5) (IH1) {IH${}_1$};
\node at (1.5,-1.5) (IH2) {IH${}_2$};

\path[->] (s) edge node[left] {$\alpha$} (t);
\path[->>] (s1) edge node[right] {$\kappa$} (t1);
\path[->>] (s2) edge node[right] {$\kappa'$} (t2);
\path[->>] (t) edge node[left] {$\rho$} (u);
\path[->>] (t2) edge node[right] {$\rho'$} (u1);
\end{tikzpicture}
}
\hfill
\subfloat[\label{fig:DIH}Pasting D and IH${}_1$ into DIH${}_1$.]{
\begin{tikzpicture}[scale=1.7]
\footnotesize
\node at (0,0) (s) {};
\node at (1,0) (s1) {};
\node at (3,0) (s2) {};

\node at (0,-1) (t) {};
\node at (1,-1) (t1) {};
\node at (3,-1) (t2) {};

\node at (0,-2) (u) {};
\node at (3,-2) (u1) {};

\path[->>](s) edge node[above] {$\tau$} (s2);
\path[->] (s) edge node[left] {$\alpha$} (t);
\path[->>] (s2) edge node[right] {$\kappa'$} (t2);
\path[->>] (t) edge node[below] {$\tau'$} (t2);

\node at (1.5,-0.5) (DIH1) {DIH${}_1$};

\phantom{\path[->>] (u) edge node[below] {$\tau''$} (u1);}
\end{tikzpicture}
}
\caption{Lemma~\ref{lem:LD_imp_D}}
\label{FIG:main}
\end{figure}

\begin{lemma}[\textnormal{similar to \cite[Theorem~3.7]{vO94}}]
\label{lem:LD_imp_D}
Let $\BB$ be a labeled ARS and $\prec$ be a transitive and well-founded
relation on the labels.
If all local peaks of~$\BB$ are decreasing, then all peaks of~$\BB$ are decreasing.
\end{lemma}
\proof
To show that all peaks are decreasing we fix a peak $(\ltoo{\tau},\ltoo{\sigma})$ and
show that this peak can be completed into a decreasing diagram.
The proof is by well-founded induction on
 $\pex$ 
and there only is the step case.
The interesting situation is when neither $\tau$ nor $\sigma$ are empty,
i.e., (using Lemma~\ref{lem:seq}(1) we obtain)
${\ltoo{\tau}} = {{\lto{\beta}}\cdot{\ltoo{\upsilon}}}$ and
${\ltoo{\sigma}} = {{\lto{\alpha}}\cdot{\ltoo{\rho}}}$ (see Figure~\ref{fig:main}).
Hence $(\lto{\beta},\lto{\alpha})$ is a local peak and from the assumption we obtain
a decreasing diagram with joining sequences $\ltoo{\kappa}$ and $\ltoo{\mu}$.
We obtain that $(\ltoo{\upsilon},\ltoo{\kappa})$ is a peak and want to show that the measure of this peak is
smaller than that of $(\ltoo{\tau},\ltoo{\sigma})$ (to apply the induction hypothesis).
Since $\beta$ is not empty with Lemma~\ref{lemma3_6} we establish
that $|(\ltoo{\upsilon},\ltoo{\kappa})|$ is smaller than $|(\ltoo{\tau},\lto{\alpha})|$ and from $|\alpha| \muleq |\sigma|$%
\footnote{\label{foot}This step is not mentioned in~\cite{vO94,vO94t} but hinted at in~\cite{vO08a}.}
we obtain the desired result.
Now, the induction hypothesis yields that IH${}_1$ is a decreasing diagram.
Concatenating (using Lemma~\ref{lem:seq}(2)) $\ltoo{\mu}$ and $\ltoo{\upsilon'}$ into a sequence $\ltoo{\tau'}$,
using Lemma~\ref{lemma3_5_DD} we can paste the diagrams D and IH${}_1$ into 
a decreasing diagram (DIH${}_1$, see Figure~\ref{fig:DIH}). 
The peak $(\ltoo{\tau'},\ltoo{\rho})$ is smaller than the peak $(\ltoo{\tau},\ltoo{\sigma})$ by a mirrored version 
of Lemma~\ref{lemma3_6} and hence the induction hypothesis yields the decreasing
diagram IH${}_2$.
Finally, a mirrored version of Lemma~\ref{lemma3_5_DD} pastes
 DIH${}_1$ and IH$_{2}$ into a decreasing diagram.
\Qed

We define local decreasingness for ARSs.

\begin{definition}[\textnormal{\cite[Definition~3.8]{vO94}}]
\label{def:ld}
An ARS~$\AA$ is \emph{locally decreasing} if there exists a 
transitive and well-founded relation $\prec$ on the labels such that all
local peaks are decreasing for (a labeled version of)~$\AA$.
\end{definition}

Finally we arrive at the main result for soundness: 

\begin{corollary}[\textnormal{\cite[Corollary~3.9]{vO94}}]
\label{cor:D_imp_CR}
A locally decreasing ARS is confluent.
\end{corollary}
\proof
From local decreasingness we get a transitive and well-founded relation $\prec$
such that all local peaks are decreasing in a labeled version of the ARS.
Lemma~\ref{lem:LD_imp_D} yields that all peaks are decreasing. The result follows
by dropping labels from the labeled rewrite sequences.
\Qed

\section{Formalization of the Conversion Version}
\label{for:cv}

In this section we give a formal proof for the main result underlying that 
local decreasingness with respect to conversions (see~\cite{vO08a}) implies confluence.
To this end we formally introduce (labeled) conversions, similarly to labeled
rewrite sequences. For each object $a$ there is the empty conversion
$a \lconvv{[]} a$ (also just written $a$)
and if $a \lto{\alpha} b$ ($a \lfrom{\alpha} b$) is a labeled rewrite step and
$b \lconvv{\sigma} c$ is a conversion then
$a \lto{\alpha} b \lconvv{\sigma} c$ ($a \lfrom{\alpha} b \lconvv{\sigma} c$)
is a conversion (often written $a \lconvv{\alpha\sigma} c$). 
For conversions we prove similar properties as for sequences (see Lemma~\ref{lem:seq}).
In addition we establish that 
mirroring a conversion again yields a conversion (with the same set of labels)
and that every sequence is a conversion.
\begin{lemma}
\label{lem:conv}
Let $a_1 \lconv{\alpha_1} \cdots \lconv{\alpha_{n-1}} a_n$ and 
$b_1 \lconv{\beta_1} \cdots \lconv{\beta_{m-1}} b_m$
be conversions. 
\begin{enumerate}
\item
Then
$a_1 \lconv{\alpha_1} \cdots \lconv{\alpha_{i-1}} a_i$ and
$a_i \lconv{\alpha_i} \cdots \lconv{\alpha_{n-1}} a_n$ are conversions
for any $1\leqslant i\leqslant n$.
\item
If $a_n = b_1$ then
$a_1 \lconv{\alpha_1} \cdots \lconv{\alpha_{n-1}} a_n = 
 b_1 \lconv{\beta_1} \cdots \lconv{\beta_{m-1}} b_m$
is a conversion.
\item
Then
$a_n \lconv{\alpha_{n-1}} 
 \cdots 
\lconv{\alpha_{1}} a_1$
is a conversion and $\{\alpha_1, 
 \ldots, 
 \alpha_n\} = \{\alpha_n,
\ldots, 
\alpha_1\}$.
\item
If $c_1 \lto{\gamma_1} \cdots \lto{\gamma_{n-1}} c_n$ is a sequence
then $c_1 \lconv{\gamma_1} \cdots \lconv{\gamma_{n-1}} c_n$ is a conversion.
\end{enumerate}
\end{lemma}
\proof
Items (1)-(3) are proved by induction on the first conversion, item (4) is 
proved by induction on the sequence.
\Qed

We will also use the following easy lemma being a direct consequence of
Definition~\ref{def:multiset}.

\begin{lemma}
\label{lem:mul_eq_ds}
If $M \muleq N$ and $N \subseteq \ds{S}$ then $M \subseteq \ds{S}$.
\qed
\end{lemma}

\begin{figure}
\subfloat[Local decreasingness wrt.\ conversions.\label{fig:ldc}]{
\begin{tikzpicture}[xscale=0.3,yscale=0.4]
\node at (0,10) (0) {};

\node at (-10,0) (1) {};
\node at (-4,0) (2) {};
\node at (4,0)  (3) {};
\node at (10,0) (4) {};

\node at (-1.5,-3) (6) {};
\node at (1.5,-3)  (7) {};

\path[->]  (0) edge node[above left] {$\alpha$} (1);
\path[->]  (0) edge node[above right] {$\beta$} (4);

\path[<<->>]  (1) edge node[below] {$\ds{\alpha}$} (2);
\path[->]     (2) edge node[xshift=-3mm] {$\beta$} node[xshift=3mm] {$=$} (6);

\path[<<->>]  (6) edge node[below] {$\ds{\alpha\beta}$} (7);

\path[<<->>]    (4) edge node[below] {$\ds{\beta}$} (3);
\path[->]     (3) edge node[xshift=3mm] {$\alpha$} node[xshift=-3mm] {$=$} (7);

\end{tikzpicture}
}
\hspace{-0.65cm}
\subfloat[Closing the conversion into a valley.\label{fig:ldc2}]{
\begin{tikzpicture}[xscale=0.4,yscale=0.5]
\footnotesize

\node at (-10,0) (1) {};
\node at (-4,0) (2) {};
\node at (4,0)  (3) {};
\node at (10,0) (4) {};

\node at (-7,-1.5) (a) {\eqref{lem:key1}};
\node at (7,-1.5) (b) {\eqref{lem:key1}};

\node at (-7,-3) (5) {};
\node at (-1,-3) (6) {};
\node at (1,-3)  (7) {};
\node at (7,-3)  (8) {};

\node at (-4,-3) (c) {\eqref{lem:key2}};
\node at (4,-3) (d) {\eqref{lem:key2}};

\node at (-6,-4) (9) {};
\node at (6,-4)  (10){};

\node at (-5,-5) (11) {};
\node at (5,-5)  (12) {};

\node at (-4,-6) (13) {};
\node at (4,-6)  (14) {};

\node at (0,-6) (d) {\eqref{lem:key1}};

\node at (0,-10) (15) {};

\path[<<->>]  (1) edge node[below] {$\ds{\alpha}$} (2);
\path[->>]    (1) edge node[xshift=-3mm] {$\ds{\alpha}$} (5);
\path[->>]    (2) edge node[xshift=3mm] {$\ds{\alpha}$} (5);
\path[->]     (2) edge node[xshift=-3mm] {$\beta$} node[xshift=3mm] {$=$}(6);
\path[->>]    (5) edge node[left] {$\ds{\alpha}$} (9);
\path[->>]    (6) edge node[right,xshift=1mm] {$\ds{\alpha\beta}$} (13);
\path[->]     (9) edge node[left,xshift=-0mm,yshift=-1mm] {$\beta$} node[xshift=3mm] {$=$} (11);
\path[->>]    (11) edge node[left,xshift=-0mm,yshift=-1mm] {$\ds{\alpha\beta}$} (13);
\path[->>]    (13) edge node[left,xshift=-0mm,yshift=-1mm] {$\ds{\alpha\beta}$} (15);

\path[<<->>]  (6) edge node[below] {$\ds{\alpha\beta}$} (7);

\path[<<->>]    (4) edge node[below] {$\ds{\beta}$} (3);
\path[->>]    (4) edge node[xshift=3mm] {$\ds{\beta}$} (8);
\path[->]     (3) edge node[xshift=3mm] {$\alpha$} node[xshift=-3mm] {$=$} (7);
\path[->>]  (3) edge node[xshift=-3mm] {$\ds{\beta}$} (8);
\path[->>]    (8) edge node[right,xshift=0mm,yshift=-1mm] {$\ds{\beta}$} (10);
\path[->>]    (7) edge node[left,xshift=-1mm] {$\ds{\alpha\beta}$} (14);
\path[->]     (10) edge node[right,xshift=-0mm,yshift=-1mm] {$\alpha$} node[xshift=-3mm] {$=$} (12);
\path[->>]    (12) edge node[right,xshift=-0mm,yshift=-1mm] {$\ds{\alpha\beta}$} (14);
\path[->>]    (14) edge node[right,xshift=-0mm,yshift=-1mm] {$\ds{\alpha\beta}$} (15);

\end{tikzpicture}
}
\caption{Conversion version of decreasing diagrams.}
\end{figure}
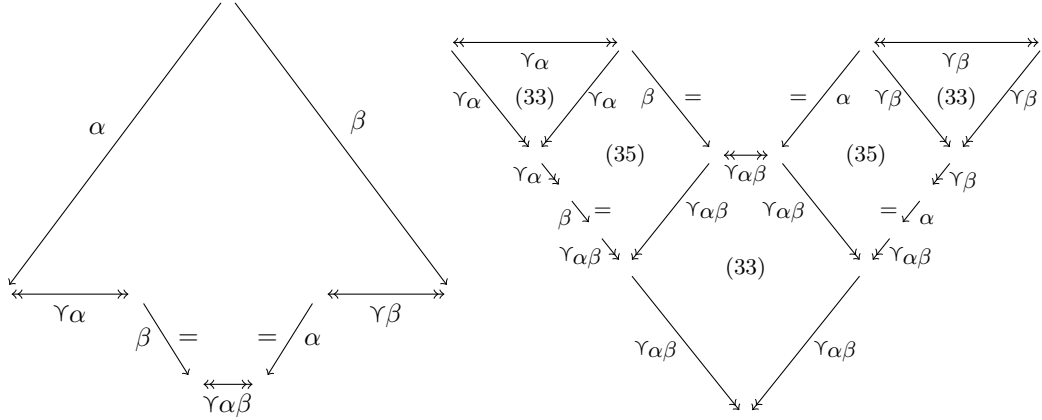

The following result (stated as observation in \cite{vO08a})
follows from Lemma~\ref{lem:mul_eq_ds}.
\begin{lemma}
\label{lem:DD_ds}
If $(\ltoo{\tau},\ltoo{\sigma},\ltoo{\sigma'},\ltoo{\tau'})$ is a decreasing diagram
and $|(\ltoo{\tau},\ltoo{\sigma})| \subseteq \ds{M}$
then also $|(\ltoo{\sigma'},\ltoo{\tau'})| \subseteq \ds{M}$.
\qed
\end{lemma}

A local peak $(\lto{\beta},\lto{\alpha})$ is \emph{decreasing with respect to conversions}%
\footnote{Please note the asymmetry to the definition of local decreasingness
(Definition~\ref{def:pd}).}
if there exist conversions such that the constraints from 
Figure~\ref{fig:ldc} are satisfied. Now we can state the main result underlying 
soundness of the conversion version of decreasing diagrams.
\begin{lemma}
\label{lem:main:conv}
Let~$\BB$ be a labeled ARS and $\prec$ be a transitive and well-founded relation 
on the labels. If all local peaks of~$\BB$ are decreasing with respect to
conversions, then all peaks of~$\BB$ are decreasing (with respect to valleys).
\end{lemma}
\proof
Similar to~\cite{vO08a} we follow the proof of the valley version 
(see Lemma~\ref{lem:LD_imp_D}). In contrast to Lemma~\ref{lem:LD_imp_D} we do not get
decreasingness of the local peak $(\lto{\beta},\lto{\alpha})$ 
(in Figure~\ref{fig:main}) by assumption.
Instead our assumption yields local decreasingness with respect to conversions, i.e., as
depicted in Figure~\ref{fig:ldc}. We close the conversion into a
valley as outlined in Figure~\ref{fig:ldc2}. To this end we use
Lemmata~\ref{lem:key1} and~\ref{lem:key2} (see below) and conclude the valleys
as shown in Figure~\ref{fig:ldc2}. Note that for the final application of
Lemma~\ref{lem:key1} we apply Lemma~\ref{lem:conv} first, to combine the sequences
and conversions into a single conversion.
Lemma~\ref{proposition3_4} (lifted to rewriting sequences) then shows
decreasingness of the diagram.
\Qed

The main structure of our proof follows the one from~\cite{vO08a}. However, there
the proofs of two key results are sketchy and informal. We identified the statements
as Lemmata~\ref{lem:key1} and~\ref{lem:key2} and provide formal proofs. 
Note that to establish these properties we can use the induction
hypothesis (from the proof of Lemma~\ref{lem:main:conv}), e.g., peaks whose measure
is smaller than $|(\lto{\beta},\lto{\alpha})|$ can be completed into a decreasing diagram.

\begin{figure}
\subfloat[Lemma~\ref{lem:key1} (case $\to$).\label{fig:key1a}]{
\begin{tikzpicture}
\footnotesize
\node at (0,0) (0)    {};
\node at (1,-1) (1)   {};
\node at (3,-1) (2)   {};
\node at (2,-2) (3)   {};
\node at (2,-1.4) (4) {IH};
\node at (0,-3) (7)    {};

\path[->]    (0) edge node[below left] {$\ds{M}$} (1);
\path[<<->>] (1) edge node[above] {$\ds{M}$} (2);
\path[->>]   (1) edge node[below left] {$\ds{M}$} (3);
\path[->>]   (2) edge node[below right] {$\ds{M}$} (3);
\end{tikzpicture}
}
\subfloat[Lemma~\ref{lem:key1} (case $\from$).\label{fig:key1b}]{
\begin{tikzpicture}
\footnotesize
\node at (0,-2) (0)   {};
\node at (1,-1) (1)   {};
\node at (3,-1) (2)   {};
\node at (2,-2) (3)   {};
\node at (1,-3) (5)   {};
\node at (2,-1.4) (4) {IH};
\node at (1,-2) (6)   {D};

\path[->]    (1) edge node[above left] {$\ds{M}$} (0);
\path[<<->>] (1) edge node[above] {$\ds{M}$} (2);
\path[->>]   (1) edge node[above right] {} (3);
\path[->>]   (2) edge node[below right] {$\ds{M}$} (3);
\path[->>]   (0) edge node[below left] {$\ds{M}$} (5);
\path[->>]   (3) edge node[below right] {$\ds{M}$} (5);
\end{tikzpicture}
}
\subfloat[Lemma~\ref{lem:key2}.\label{fig:key2a}]{
\begin{tikzpicture}[xscale=0.5,yscale=0.5]
\footnotesize
\node at (0,0) (0)   {};
\node at (-3,-3) (1) {};
\node at (3,-3) (2)  {};
\node at (-2,-4) (3) {};
\node at (-1,-5) (4) {};
\node at (0,-6) (5)  {};
\node at (0,-3) (d)  {D};

\path[->]    (0) edge node[above left] {$\alpha$}            node[below right] {$\sigma$}
 node[very near end,rotate=45,below]{$=$} (1);
\path[->>]   (0) edge node[above right] {$\ds{\beta}$}  node[below left] {$\tau$}(2);
\path[->>]   (1) edge                                   node[above right] {$\tau'$}(5);
\path[->>]   (2) edge                                   node[above left] {$\sigma'$}(5);

\end{tikzpicture}
}
\subfloat[Lemma~\ref{lem:key2}.\label{fig:key2b}]{
\begin{tikzpicture}[xscale=0.5,yscale=0.5]
\footnotesize
\node at (0,0) (0)   {};
\node at (-3,-3) (2) {};
\node at (3,-3) (1)  {};
\node at (2,-4) (3) {};
\node at (1,-5) (4) {};
\node at (0,-6) (5)  {};

\path[->>]   (0) edge node[above right]{$\ds{\beta}$}        node[below left] {$\tau$}(1);
\path[->]    (0) edge node[above left] {$\alpha$}            node[below right] {$\sigma$}
 node[very near end,rotate=45,below]{$=$} (2);
\path[->>]   (1) edge node[below right] {$\ds{\beta}$}       node[above left] {$\sigma_1$}(3);
\path[->>]   (2) edge node[below left]{$\ds{\alpha\beta}$}   node[above right] {$\tau'$}(5);
\path[->>]   (4) edge node[below right] {$\ds{\alpha\beta}$} node[above left] {$\sigma_3$}(5);

\path[->]    (3) edge node[below right] {$\alpha$}           node[above left] {$\sigma_2$}
 node[very near end,rotate=45,above,yshift=-0.5mm]{$=$} (4);
 (4);
\end{tikzpicture}
}
\caption{Lemmata~\ref{lem:key1} and~\ref{lem:key2}.}
\end{figure}

\begin{lemma}
\label{lem:key1}
Let all peaks smaller than $|(\lto{\beta},\lto{\alpha})|$ have a decreasing diagram.
Then for any $M$ with $M \muleq \{\#\alpha,\beta\#\}$ we have
${\lconvv{\ds{M}}} \subseteq {\ltoo{\ds{M}} \cdot \lfromm{\ds{M}}}$.
\end{lemma}
\proof
By induction on the conversion $\lconvv{\ds{M}}$. The base case is trivial. In
the step case we have $\lconv{\ds{M}} \cdot \lconvv{\ds{M}}$. The induction hypothesis
yields $\lconv{\ds{M}} \cdot \ltoo{\ds{M}} \cdot \lfromm{\ds{M}}$. We consider two cases.
If the first step is from left to right, i.e., $\lto{\ds{M}}$ then the result follows
from Lemma~\ref{lem:conv}(2) (see Figure~\ref{fig:key1a}).
In the other case we have $\lfrom{\ds{M}} \cdot \ltoo{\ds{M}} \cdot \lfromm{\ds{M}}$.
Since the peak $\lfrom{\ds{M}} \cdot \ltoo{\ds{M}}$ has a smaller measure
than 
$(\lto{\beta},\lto{\alpha})$
it can be completed into a decreasing diagram
and Lemma~\ref{lem:DD_ds} in combination with Lemma~\ref{lem:conv}(2)
yields the result (see Figure~\ref{fig:key1b}).
\Qed

To show the second key result we establish a useful decomposition result on sequences.
\begin{lemma}
\label{lem:seq_dec}
Let $\ltoo{\sigma}$ be a sequence and $\sigma = \sigma_1\sigma_2$.
Then there are sequences $\ltoo{\sigma_1}$ and $\ltoo{\sigma_2}$ such
that ${\ltoo{\sigma}} = {\ltoo{\sigma_1} \cdot \ltoo{\sigma_2}}$.
\end{lemma}
\proof
By induction on the sequence $\ltoo{\sigma}$.
\Qed

Below $\ltoe{\alpha}$ stands for $\lto{\alpha}$ (one step) or $\ltoo{[]}$ (zero steps).
Please note the similarity of the following result to the explicit characterization
of local decreasingness (cf.\ Figure~\ref{fig:ld'}).

\begin{lemma}
\label{lem:key2}
Let all peaks smaller than $|(\lto{\beta},\lto{\alpha})|$ have a decreasing diagram.
Then the peak $(\ltoo{\ds{\beta}},\ltoe{\alpha})$ can be closed by
\({
\ltoo{\ds{\alpha\beta}}
\cdot 
\lfromm{\ds{\alpha\beta}}\cdot \lfrome{\alpha}\cdot \lfromm{\ds{\beta}}
}\) (see Figure~\ref{fig:key2b}).
\end{lemma}
\proof
Since $|(\ltoo{\ds{\beta}},\ltoe{\alpha})|$ is smaller than
$|(\lto{\beta},\lto{\alpha})|$, it can be completed into a decreasing diagram
$(\ltoo{\tau},\ltoo{\sigma},\ltoo{\sigma'},\ltoo{\tau'})$ (see Figure~\ref{fig:key2a}).
First we show $\tau' \subseteq \ds{\alpha\beta}$. From decreasingness
and Lemma~\ref{lem:decreasing} we get $|\tau'| \diffs \ds{\sigma} \muleq |\tau|$.
The assumption $\tau \subseteq \ds{\beta}$ and Lemma~\ref{proposition3_4_aux} yields
$|\tau| \subseteq \ds{\beta}$.
Using Lemma~\ref{lem:mul_eq_ds}
we obtain $|\tau'| \diffs \ds{\sigma} \subseteq \ds{\beta}$,
i.e.\ $|\tau'| \subseteq \ds{\beta} \cup \ds{\sigma}$.
The assumption $\sigma \subseteq \alpha$ yields $\ds{\sigma} \subseteq \ds{\alpha}$
and hence we conclude by Lemmata~\ref{lemma2_6}(1) and~\ref{proposition3_4_inv_aux}(2).

Next we show that $\ltoo{\sigma'}$ can be decomposed into 
$\ltoo{\sigma_1}$, $\ltoe{\sigma_2}$, and $\ltoo{\sigma_3}$ with
$\sigma_1 \subseteq \ds{\beta}$, $\sigma_2 \subseteq \{\alpha\}$, $\length{\sigma_2} \leq 1$, 
and $\sigma_3 \subseteq \ds{\alpha\beta}$. To this end we first
observe that Lemma~\ref{proposition3_4_inv_step} also holds if $\beta$ is not a
single label but a sequence (here $\tau$). Then from decreasingness we obtain
$\sigma' = \sigma_1\sigma_2\sigma_3 \land \sigma_1 \subseteq \ds{\tau}
\land \length \sigma_2 \leq 1 \land \sigma_2 \subseteq \{\alpha\} \land
\sigma_3 \subseteq \ds{\alpha\sigma}$. Lemma~\ref{lem:seq_dec} lifts the
decomposition of labels to a decomposition of sequences and we can conclude.
\Qed

An ARS~$\AA$ is \emph{locally decreasing with respect to conversions} if there exists a
transitive and well-founded relation $\prec$ on the labels such that all local
peaks are decreasing with respect to conversions for (a labeled version of)~$\AA$.
Finally we arrive at the main result for soundness: 

\begin{corollary}[\textnormal{\cite[Theorem~3]{vO08a}}]
\label{cor:LDC_imp_CR}
A locally decreasing with respect to conversions ARS is confluent.
\qed
\end{corollary}

\section{Meanderings}
\label{mea:main}

In this section we discuss differences between our formalization and
(proofs from)~\cite{vO94,vO08a}. 

Within \Isabelle (\thy{Abstract\_Rewriting.thy}) an ARS is a binary
relation while in~\cite{vO94} the ARS also contains the domain of the 
relation. A similar statement holds for labeled ARSs.

\emph{General multisets} are used in \cite{vO94}, which can represent sets and
finite multisets in one go wheres our formalization clearly separates the two
concepts. The reason is purely practical, i.e., the \Isabelle library already
contains the dedicated theories \thy{Set.thy} and \thy{Multiset.thy}.
The only (negligible) 
disadvantage we have experienced from this design choice is the need for 
multiple definitions of the down-set (for lists, sets, and multisets)
and for Lemma~\ref{lemma2_6}(1). On the other hand, this saved us from 
formalizing \emph{general multisets}, which we anticipate
as a significant endeavour on its own.
Moreover,~\cite{vO94} uses a different multiset extension 
than \thy{Multiset.thy}. The latter defines the multiset extension 
as the transitive closure of the ``one-step'' multiset extension.
\begin{definition}
The \emph{one-step multiset extension} (denoted by $\multone$) of~$\prec$ is
defined by 
\[
\text{$M \multone N$ if $\exists$ $a$ $I$ $K$. $M = I + K$, $N = I + \{\#a\#\}$, 
 $\forall\ b \in K$. $b \prec a$}
\]
and the \emph{multiset extension} of $\prec$ (denoted by $\mult$) is the transitive closure
of $\multone$.
\end{definition}

Based on the results in \thy{Multiset.thy} and Definition~\ref{def:multiset}(1) we
have proven these two definitions equivalent for any transitive base relation.
\begin{lemma}
\label{lem:mult}
If $\prec$ is transitive then $\mult$ and $\mul$ coincide.
\qed
\end{lemma}

Moreover we proved the claim in Definition~\ref{def:multiset}.
\begin{lemma}
\label{lem:multeq}
We have that $\muleq$ is the reflexive closure of $\mul$.
\qed
\end{lemma}
\proof
First we show the inclusion from left to right.
Let $M \muleq N$. If $J = \{\#\}$ then $M = N$ and the result follows.
If $J \not= \{\#\}$ then $M \mul N$ and we are done.

For the reverse inclusion let $(M,N)$ be in the reflexive closure of $\mul$.
If $M = N$ then we finish with $I = M$, $K = J = \{\#\}$. In the other
case we get suitable $I$, $J$, and $K$ from the definition of $\mul$.
\Qed

Our formalization is first performed for sequences (of labels) and then lifted to
labeled rewrite sequences (conversions), a step which is left implicit in~\cite{vO94}.
After introducing labeled rewriting, we proved useful results in \Isabelle 
(Lemmata~\ref{lem:seq} and~\ref{lem:conv}).

In addition to the algebraic proof of Lemma~\ref{lemma2_6}(3) from~\cite{vO94}
our formalization contains an alternative one.
Our proof of~Lemma~\ref{lemma3_2}(1) differs from the informal one
in~\cite{vO94}. 
Also the formal proof of Lemma~\ref{proposition3_4} differs from the sketch given
for \cite[Proposition~3.4]{vO94}, requiring auxiliary results 
(Lemmata~\ref{proposition3_4_aux} and~\ref{proposition3_4_inv_aux}).

There are some (tiny) differences between~\cite[Theorem~3.7]{vO94} and
Lemma~\ref{lem:LD_imp_D}. In \cite{vO94} a measure on \emph{diagrams} is used.
However, since the closing/joining steps of the diagram are just
obtained by the induction hypothesis the measure must be on \emph{peaks} (which
is used in~\cite{vO08a}).
Moreover, since in either case the measure is a multiset it is hard to relate
arbitrary multisets to a peak. Hence we lifted the order on labels $\prec$ to peaks
$\pex$ (Section~\ref{for:mr}) and used well-founded induction on this order.
In the formalization of Lemma~\ref{lem:LD_imp_D} (Footnote~\ref{foot})
we identified a necessary step to apply the induction hypothesis.
Another aspect where our formalization deviates from~\cite{vO94} is that
the original work uses families of labeled ARSs whereas our formalization
considers a single labeled ARS only. Hence  
\cite[Theorem~3.7]{vO94} states the main result on families of 
ARSs whereas our Lemma~\ref{lem:LD_imp_D} makes a statement
about a single 
ARS.

Concerning~\cite{vO94} our formal proofs for the alternative formulation of local
decreasingness (Lemma~\ref{proposition3_4}) differs from the one in~\cite{vO94,vO94t}.
While this alternative formulation of local decreasingness was not needed to
obtain the main result underlying the valley version
(\cite[Main~Theorem~3.7]{vO94}, i.e., Lemma~\ref{lem:LD_imp_D}), it was 
(in a generalized formulation) essential for the main result underlying the 
conversion version (\cite[Theorem~3]{vO08a}, i.e., Lemma~\ref{lem:main:conv}).
Furthermore we gave formal proofs for two (informal) key
observations made in the proof of~\cite[Theorem~3]{vO08a}, resulting in
Lemmata~\ref{lem:key1} and~\ref{lem:key2}. Especially the latter has a non-trivial
formal proof, since the induction hypothesis yields decreasingness 
(see Figure~\ref{fig:key2a}) but not the desired decomposition of the joining
sequences (see Figure~\ref{fig:key2b}), in contrast to what 
the proof in \cite{vO08a} conveys.

\section{Conclusion}
\label{con:main}
 
In this paper we have described a formalization of decreasing diagrams
in the theorem prover~\Isabelle following the original proofs 
from~\cite{vO94,vO08a}. In Sections~\ref{for:ld} and \ref{for:rew} our
formal proofs deviate from the either informal or implicit ones in~\cite{vO94}
and we also elaborate on Lemma~\ref{lem:key2}, a result which is implicitly 
used in~\cite{vO08a}.
To show the applicability of our formalization we performed a mechanical proof of
Newman's lemma using decreasing diagrams (following~\cite[Corollary~4.4]{vO94}).
Our formalization has few dependencies on existing theories. 
From \thy{Abstract\_Rewriting.thy} we employ some properties for unlabeled abstract
rewriting (and the definition of confluence).
The theory \thy{Multiset.thy} provides standard multiset operations and a
well-foundedness proof of the multiset extension of a well-founded relation. 
Note that some of our results on multisets (a formalized proof 
of~\cite[Lemma~2.6(3)]{vO94}, i.e., Lemma~\ref{lemma2_6}(3))
might be of interest for a larger community.

In~\cite{B97} a ``point version'' of decreasing diagrams is introduced, where 
objects are labeled instead of steps. It is unknown if the point version is 
equivalent to the standard one.
Parts of~\cite{B97} have been formalized in \COQ but 29 axioms are assumed, i.e., 
not proven in the theorem prover. Furthermore the more useful alternative
representation of local decreasingness (Lemma~\ref{proposition3_4}) is not
considered in~\cite{B97}. The same holds for the conversion version.
Hence~\cite{B97} is only a \emph{partial} formalization and essentially
different from ours.

We anticipate that our contribution paves the way for future work in
several directions. One possibility is the
formalization of confluence results that can be proven with
decreasing diagrams (e.g.\ Toyama's theorem~\cite{T87}).
The benefit might be two-fold. On the one hand side the proof by
decreasing diagrams might be easier to formalize and furthermore
proofs by decreasing diagrams are constructive, cf.\ \cite{vO08b}.
Another idea would be the certification of confluence proofs 
(based on decreasing diagrams) given by automated confluence provers.%
\footnote{Certification is already established in the termination community
where it has shown tools as well as termination criteria unsound.}
Both aims require to lift our formalization 
from abstract rewriting to term rewriting, which is a natural idea for future 
work.
\drop{We stress that the
\emph{Isabelle Formalization of Rewriting} (\ISAFOR~\cite{CETA}) already
contains notions such as critical pairs, which will ease this job. \ISAFOR
has been developed to formalize termination criteria for rewriting and 
also offers the opportunity to check concrete termination proofs given by
automated termination tools. A dedicated category is present in the
international competition of termination tools%
\footnote{\url{http://termcomp.uibk.ac.at}}
since 2007. Concerning the confluence competition,%
\footnote{\url{http://coco.nue.riec.tohoku.ac.jp/}} already in its first
edition confluence proofs due to Knuth and Bendix' criterion~\cite{KB70} and
for orthogonal systems~\cite{R73} could be certified with the help of \ISAFOR. 
These two criteria applied to 27 out of 113 confluence proofs and hence
our contribution can be seen as a first step to drastically increase the number of
certified confluence proofs.
}

\medskip\emph{Acknowledgments:}
This research is supported by FWF P22467.
We thank the anonymous reviewers, Bertram Felgenhauer, Nao Hirokawa, and
Aart Middeldorp for helpful comments.
Bertram Felgenhauer contributed an initial proof of Lemma~\ref{lemma2_6}(3)
and located the formalization of~\cite{B97}.

\clearpage
\appendix

\section{Isabelle Definitions}
\label{def:main}
Definition~\ref{def:multiset} can easily be mimicked in \Isabelle
(here \isa{ds}/\isa{dm}/\isa{dl} defines the down-set for a set/multiset/list):%
\footnote{For readability of subsequent definitions we denote $\prec$ by \isa{r}
within code listings.}

\begin{isabelle}
definition ds :: "'a rel $\Rightarrow$ 'a set $\Rightarrow$ 'a set"
 where "ds r S = {y . $\exists$x $\in$ S. (y,x) $\in$ r}"

definition dm :: "'a rel $\Rightarrow$ 'a multiset $\Rightarrow$ 'a set" 
 where "dm r M = ds r (set_of M)"

definition dl :: "'a rel $\Rightarrow$ 'a list $\Rightarrow$ 'a set" 
 where "dl r $\sigma$ = ds r (set $\sigma$)"

definition mul :: "'a rel $\Rightarrow$ 'a multiset rel" where
  "mul r = {(M,N).$\exists$I J K. M = I + K $\land$ N = I + J $\land$ set_of K $\subseteq$ dm r J $\land$ J $\not=$ {#}}"

definition mul_eq :: "'a rel $\Rightarrow$ 'a multiset rel" where
  "mul_eq r = {(M,N).$\exists$I J K. M = I + K $\land$ N = I + J $\land$ set_of K $\subseteq$ dm r J}"
\end{isabelle}

Since the lexicographic maximum measure depends on the base order $\prec$ on labels, 
in \Isabelle Definition~\ref{def:lexmax} amounts to:
\begin{isabelle}
fun lexmax :: "'a rel $\Rightarrow$ 'a list $\Rightarrow$ 'a multiset" ("(_|_|)") where 
   "r|[]| = {#}"
 | "r|$\alpha$#$\sigma$| =  {#$\alpha$#} + (r|$\sigma$| -s ds r {$\alpha$})" 
\end{isabelle}

Definition~\ref{def:decreasing} has a one-to-one correspondence in \Isabelle:
\begin{isabelle}
definition decreasing::"'a rel $\Rightarrow$ 'a list $\Rightarrow$ 'a list $\Rightarrow$ 'a list $\Rightarrow$ 'a list $\Rightarrow$ bool" 
 where "decreasing r $\tau$ $\sigma$ $\sigma$' $\tau$' = ((r|$\sigma$@$\tau$'|, r|$\tau$| + r|$\sigma$| ) $\in$ mult_eq r 
                                $\land$ (r|$\tau$@$\sigma$'|, r|$\tau$| + r|$\sigma$| ) $\in$ mult_eq r)"
\end{isabelle}

In the sequel objects will have type \isa{'a} and labels will have type \isa{'b}.
A labeled rewrite step carries the label between its two objects and is hence of type
\isa{'a $\times$ 'b $\times$ 'a}. A labeled ARS is a set of labeled rewrite steps. 
\begin{isabelle}
type_synonym ('a,'b) lars = "('a$\times$'b$\times$'a) set"
\end{isabelle}

The sequence from Example~\ref{ex:seq} is represented as
\isa{(a,[($\alpha$,b),($\beta$,c)])} in \Isabelle.
Empty sequences consist of at least an object, i.e.,
the empty sequence starting from $a$ is \isa{(a,[])}.
\begin{isabelle}
type_synonym ('a,'b) seq = "('a$\times$('b$\times$'a) list)"

inductive_set seq :: "('a,'b) lars $\Rightarrow$ ('a,'b) seq set" for $\BB$ where 
   "(a,[]) $\in$ seq $\BB$" 
 | "(a,$\alpha$,b) $\in$ $\BB$ $\Longrightarrow$ (b,ss) $\in$ seq $\BB$ $\Longrightarrow$  (a,($\alpha$,b) # ss) $\in$ seq $\BB$"
\end{isabelle}

We define \isa{lst}, which computes the \emph{last} element of 
a rewrite sequence. 

\begin{isabelle}
definition lst :: "('a,'b) seq $\Rightarrow$ 'a"
 where "lst ss = (if snd ss = [] then fst ss else snd (last (snd ss)))"
\end{isabelle}

From now on we use $\tau$, $\sigma$, etc.\ also to denote (labeled rewrite) sequences in 
\Isabelle. The type information clarifies if labels or rewrite sequences are meant.
We mimic Definition~\ref{def:diagram} in \Isabelle.

\begin{isabelle}
definition diagram :: 
 "('a,'b) lars  $\Rightarrow$ ('a,'b) seq $\times$ ('a,'b) seq $\times$ ('a,'b) seq $\times$ ('a,'b) seq $\Rightarrow$ bool"
 where "diagram $\BB$ d = (let ($\tau$,$\sigma$,$\sigma$',$\tau$') = d in {$\sigma$,$\tau$,$\sigma$',$\tau$'} $\subseteq$ seq $\BB$ $\land$
   fst $\sigma$ = fst $\tau$ $\land$ lst $\sigma$ = fst $\tau$' $\land$ lst $\tau$ = fst $\sigma$' $\land$ lst $\sigma$' = lst $\tau$')"
\end{isabelle}

Next we introduce a function \isa{labels}, which extracts the labels of 
a sequence, e.g., \isa{labels}$(a \lto{\alpha} b \ltoo {\beta} c) = [\alpha,\beta]$.
With the help of this function we can define 
a predicate \isa{DD}, which holds if a quadruple of sequences
forms a decreasing diagram.

\begin{isabelle}
definition labels ::
 "('a,'b) seq $\Rightarrow$ ('a,'b) seq $\times$ ('a,'b) seq $\times$ ('a,'b) seq $\times$ ('a,'b) seq' b $\Rightarrow$ list"
 where "labels ss = map fst (snd ss)"

definition DD :: "('a,'b) lars $\Rightarrow$ 'b rel $\Rightarrow$ $\Rightarrow$ bool"
 where "DD $\BB$ r d = (let ($\tau$,$\sigma$,$\sigma$',$\tau$') = d in 
  diagram $\BB$ d $\land$ decreasing r (labels $\tau$) (labels $\sigma$) (labels $\sigma$') (labels $\tau$'))"
\end{isabelle}

Definition~\ref{def:measure} reads as follows:
\begin{isabelle}
definition measure :: "'b rel $\Rightarrow$ ('a,'b) seq $\times$ ('a,'b) seq $\Rightarrow$ 'b multiset"
 where "measure r p = r|labels (fst p)| + r|labels (snd p)|"

definition pex :: "'b rel $\Rightarrow$ ('a,'b) seq $\times$ ('a,'b) seq"
 where "pex r = {(p1,p2). (measure r p1,measure r p2) $\in$ mul r}"
\end{isabelle}

Next peaks and local peaks (see Definition~\ref{def:peak}) are introduced.
\begin{isabelle}
definition peak :: "('a,'b) lars $\Rightarrow$ ('a,'b) seq $\times$ ('a,'b) seq $\Rightarrow$ bool"
 where "peak lars p = (let ($\tau$,$\sigma$) = p in {$\tau$,$\sigma$} $\subseteq$ seq lars $\land$ fst $\tau$ = fst $\sigma$)"

definition local_peak :: "('a,'b) lars $\Rightarrow$ ('a,'b) seq $\times$ ('a,'b) seq $\Rightarrow$ bool"
 where "local_peak lars p = (let ($\tau$,$\sigma$) = p in 
 peak lars p $\land$ length (snd $\tau$) = 1 $\land$ length (snd $\sigma$) = 1)"
\end{isabelle}

The following definition (corresponding to Definition~\ref{def:ld}) shows
that the labeled version of $\AA$ can be chosen freely since we only demand
the existence of a labeled version of $\AA$ satisfying decreasingness of all
local peaks.
\begin{isabelle}
definition unlabel :: "('a,'b) lars $\Rightarrow$ 'a rel"
 where "unlabel $\BB$ = {(a,c). $\exists$b. (a,b,c) $\in$ $\BB$}"

definition LD :: "'b set $\Rightarrow$ 'a rel $\Rightarrow$ bool"
 where "LD L $\AA$ = ($\exists$ r $\BB$. ($\AA$ = unlabel $\BB$) $\land$ trans r $\land$ wf r $\land$ 
 ($\forall$p. (local_peak $\BB$ p $\longrightarrow$ ($\exists$ $\sigma$' $\tau$'. (DD $\BB$ r (fst p,snd p,$\sigma$',$\tau$'))))))"
\end{isabelle}

Conversions are defined in \Isabelle as follows:
\begin{isabelle}
type_synonym ('a,'b) conv = "('a $\times$ ((bool $\times$ 'b $\times$ 'a) list))"

inductive_set conv :: "('a,'b) lars $\Rightarrow$ ('a,'b) conv set" for ars 
where "(a,[]) $\in$ conv ars" 
    | "(a,$\alpha$,b) $\in$ ars $\Longrightarrow$ (b,ss) $\in$ conv ars $\Longrightarrow$ (a,(True,$\alpha$,b) # ss) $\in$ conv ars"
    | "(b,$\alpha$,a) $\in$ ars $\Longrightarrow$ (b,ss) $\in$ conv ars $\Longrightarrow$ (a,(False,$\alpha$,b) # ss) $\in$ conv ars"
\end{isabelle}


\begin{thebibliography}{10}

\bibitem{BKO98}
Bezem, M., Klop, J., V.{~van Oostrom}:
\newblock Diagram techniques for confluence.
\newblock I\&C 141(2), 172--204 (1998)

\bibitem{B97}
Bognar, M.:
\newblock A point version of decreasing diagrams.
\newblock In: Proceedings Accolade 1996. Dutch Graduate School in Logic. pp.
  1--14 (1997).
\newblock The formalization is available from
  \url{http://web.archive.org/web/20051226052550/http://www.cs.vu.nl/~mirna/}

\bibitem{CCFPU11}
Contejean, E., Courtieu, P., Forest, J., Pons, O., Urbain, X.:
\newblock Automated certified proofs with {CiME3}.
\newblock In: Proc.\ 22nd RTA. LIPIcs, vol. 10, pp. 21--30 (2011)

\bibitem{DM79}
Dershowitz, N., Manna, Z.:
\newblock Proving termination with multiset orderings.
\newblock Comm.\ ACM 22(8), 465--476 (1979)

\bibitem{F12c}
Felgenhauer, B.:
\newblock A proof order for decreasing diagrams.
\newblock In: Proc.\ 1st IWC. pp. 7--14 (2012)

\bibitem{GAR08}
Galdino, A., Ayala-Rinc\'on, M.:
\newblock A formalization of {N}ewman's and {Y}okouchi's lemmas in a
  higher-order language.
\newblock JFR 1(1), 39--50 (2008)

\bibitem{GAR10}
Galdino, A., Ayala-Rinc{\'o}n, M.:
\newblock A formalization of the {Knuth-Bendix(-Huet)} critical pair theorem.
\newblock JAR 45(3), 301--325 (2010)

\bibitem{H94}
Huet, G.:
\newblock Residual theory in lambda-calculus: A formal development.
\newblock JFP 4(3), 371--394 (1994)

\bibitem{JvO09}
Jouannaud, J.P., {van Oostrom}, V.:
\newblock Diagrammatic confluence and completion.
\newblock In: Proc.\ 36th ICALP. LNCS, vol. 5556, pp. 212--222 (2009)

\bibitem{KOV00}
Klop, J., {van Oostrom}, V., de~Vrijer, R.:
\newblock A geometric proof of confluence by decreasing diagrams.
\newblock JLP 10(3), 437--460 (2000)

\bibitem{N01}
Nipkow, T.:
\newblock More {C}hurch-{R}osser proofs.
\newblock JAR 26(1), 51--66 (2001)

\bibitem{ISABELLE}
Nipkow, T., Paulson, L., Wenzel, M.:
\newblock Isabelle/HOL -- A Proof Assistant for Higher-Order Logic. vol. 2283
  of LNCS.
\newblock Springer (2002)

\bibitem{vO94}
{van Oostrom}, V.:
\newblock Confluence by decreasing diagrams.
\newblock TCS 126(2), 259--280 (1994)

\bibitem{vO94t}
{van Oostrom}, V.:
\newblock Confluence for Abstract and Higher-Order Rewriting.
\newblock PhD thesis, Vrije Universiteit, Amsterdam (1994)

\bibitem{vO08a}
{van Oostrom}, V.:
\newblock Confluence by decreasing diagrams -- converted.
\newblock In: Proc.\ 19th RTA. LNCS, vol. 5117, pp. 306--320 (2008)

\bibitem{vO08b}
{van Oostrom}, V.:
\newblock Modularity of confluence constructed.
\newblock In: Proc.\ 4th IJCAR. LNCS, vol. 5195, pp. 348--363 (2008)

\bibitem{vO12}
van Oostrom, V.:
\newblock Decreasing proof orders -- interpreting conversions in involutive
  monoids.
\newblock In: Proc.\ 1st IWC. pp. 1--4 (2012)

\bibitem{P92}
Pfenning, F.:
\newblock A proof of the {C}hurch-{R}osser theorem and its representation in a
  logical framework.
\newblock Technical Report CMU-CS-92-186, School of Computer Science, Carnegie
  Mellon University (1992)

\bibitem{RRAHMM02}
Ruiz-Reina, J.L., Alonso, J.A., Hidalgo, M.J., Mart\'in-Mateos, F.J.:
\newblock Formal proofs about rewriting using {ACL2}.
\newblock AMAI 36(3), 239--262 (2002)

\bibitem{S88}
Shankar, N.:
\newblock A mechanical proof of the {C}hurch-{R}osser theorem.
\newblock JACM 35(3), 475--522 (1988)

\bibitem{ST10}
Sternagel, C., Thiemann, R.:
\newblock Abstract rewriting.
\newblock AFP (2010)

\bibitem{S06}
St{\o}vring, K.:
\newblock Extending the extensional lambda calculus with surjective pairing is
  conservative.
\newblock LMCS 2(2), 14 pages (2006)

\bibitem{T95}
Takahashi, M.:
\newblock Parallel reductions in $\lambda$-calculus.
\newblock I\&C 118(1), 120--127 (1995)

\bibitem{TeReSe}
Terese:
\newblock Term Rewriting Systems. vol.~55 of Cambridge Tracts in Theoretical
  Computer Science.
\newblock Cambridge University Press (2003)

\bibitem{T12}
Thiemann, R.:
\newblock Certification of confluence proofs using {CeTA}.
\newblock In: Proc.\ 1st IWC. p.~45 (2012)

\bibitem{T87}
Toyama, Y.:
\newblock On the {C}hurch-{R}osser property for the direct sum of term
  rewriting systems.
\newblock JACM 34(1), 128--143 (1987)

\bibitem{Z13corr}
Zankl, H.:
\newblock Confluence by decreasing diagrams -- formalized.
\newblock CoRR abs/1210.1100v2, 15 pages (2013)

\end{thebibliography}
\end{document}